\documentclass[conference]{IEEEtran}
\usepackage{fixltx2e}
\usepackage{graphicx}
\usepackage{color}
\usepackage{rotating}
\usepackage{amsfonts}
\usepackage{pifont}
\usepackage{program} 
\usepackage{float}

\ifCLASSINFOpdf
\else
\fi
\begin{document}
%
\title{Capturing Sensor Data from Mobile Phones using \\Global Sensor Network Middleware}



%
\author{\IEEEauthorblockN{Charith Perera\IEEEauthorrefmark{1}\IEEEauthorrefmark{2},
 Arkady Zaslavsky\IEEEauthorrefmark{2},
Peter Christen\IEEEauthorrefmark{1}, 
 Ali Salehi\IEEEauthorrefmark{2} and
Dimitrios Georgakopoulos\IEEEauthorrefmark{2}}
\IEEEauthorblockA{\IEEEauthorrefmark{1}Research School of Computer Science, The Australian National University,Canberra, ACT 0200, Australia}
\IEEEauthorblockA{\IEEEauthorrefmark{2}CSIRO ICT Center, Canberra, ACT 0200, Australia}}


\maketitle

\begin{abstract}
Mobile phones play increasingly bigger role in our everyday lives. Today, most smart phones comprise a wide variety of sensors which can sense the physical environment. The Internet of Things vision encompasses participatory sensing which is enabled using mobile phones based sensing and reasoning. In this research, we propose and demonstrate our DAM4GSN architecture to capture sensor data using sensors built into the mobile phones. Specifically, we combine an open source sensor data stream processing engine called `Global Sensor Network (GSN)' with the Android platform to capture sensor data. To achieve this goal, we proposed and developed a prototype application that can be installed on Android devices as well as a \textit{AndroidWrapper} as a GSN middleware component. The process and the difficulty of manually connecting sensor devices to sensor data processing middleware systems are examined. We evaluated the performance of the system based on power consumption of the mobile client. \end{abstract}


%
\IEEEpeerreviewmaketitle

\section{Introduction}
\label{sec:Introduction}

Mobile phones have built-in sensors that can be used to measure different parameters such as motion, position and various environmental conditions. The sensing capability and data processing power of mobile phones have been increased during the past decade. A decade ago, there weren't any processors in mobile phones. The industry has evolved and latest mobile phones comprise 1.4 GHz dual-core processors and 1GB ram. This enables significant potential to build sensor networks using mobile phones. One of the major challenges in sensor networks is deployment of sensors. Deployment of traditional sensor network requires to bear significant cost and effort. However, sensors in mobile phones do not require a deployment. Mobile phones are already in the hands of 5.6 billion \cite{P228} people around the world. Therefore, it is cheap to use these sensors built into the mobile phones. This fact motivates us to capture sensor data using mobile phones. Without putting substantial effort into typical sensor deployment, sensors built-in to the mobile phones can be utilised to understand mobile users and the environments around them. The raw data produced by these sensors built-in the mobile phones need to be collected and processed. Sensor data stream processing engines provide a solution for this \cite{P022}.

Internet of Things (IoT) \cite{P029} is a concept that is closely related to and motivated by sensor networks. The European Union has defined the IoT vision  and explained the applications in detail in \cite{P019}. Mobile phones are a critical component in the IoT as they are capable of performing more computation than other \textit{smart objects} \cite{P041}.
%
%
%

Using mobile phones as sensors has a significant advantage over unattended wireless sensor networks. We do not need to put extra effort to power the built-in sensors in mobile phones. Mobile phones are powered (charged) by human beings regularly. Furthermore, sensors built into mobile phones can provide more coverage than static sensors.

The availability of cheap, cost effective and widely available sensors built-in to mobile phones enable a whole new range of applications across a wide variety of domains, such as healthcare, smart home and office, social networks, safety, public parks, shopping malls, environmental monitoring, transportation and logistics. In addition, it is very easy to distribute applications for mobile phones. Almost all the mobile platforms provide easy methods to download and install applications for mobile phones such as iPhone (App Store iOS), Android (Android Market), Windows (Windows Phone Marketplace) and Blackberry (Blackberry App World). This has enabled the possibility to reach millions of users very easily.

Mobile phone sensing is a fairly new field which emerged with the increasing sensing capability of modern mobiles phones. Lane et al. \cite{P217} has conducted a survey on mobile phone sensing. Mobile sensing does raise many question in term of security and privacy. These two issues are discussed in detail in \cite{P229, P230}. As mentioned earlier, sensor data collection and processing is also an important component in mobile phone sensing systems. Sensor data stream processing engines such as Global Sensor Network \cite{P167} provides basic functionalities that required by such a system.

Let's introduce an application scenario. A farmer visits his field of crops and collects sensor data from variety of different sensors deployed. The mobile phone annotates collected raw sensor data with various context information such as location, time, etc. and sends them to GSN for storage, analysis, and interpretation. Our main contribution, DAM4GSN architecture, allows GSN to capture context annotated sensor data from low-level computational devices such as mobile phones without porting GSN itself into mobile phones.

The rest of the paper is organised as follows. Section \ref{sec:Mobile Phone Sensors} presents an overview on sensing capability of the mobile phones. Section \ref{sec:Global Sensor Networks} describes the Global Sensor Network middleware in brief. A detailed description about the entire system, including client configuration, server configuration, and data formats is provided in Section \ref{sec:System Description}. The life cycle of the developed wrappers is discussed in Section \ref{sec:GSN Wrapper's Life Cycle}. Section \ref{sec:Android Wrapper} provides a detailed explanation of wrapper development. Finally, Section \ref{sec:Evaluation} presents the performance evaluation of the proposed system.


\section{Mobile Phone Sensors}
\label{sec:Mobile Phone Sensors}

\begin{figure}[!ht]
 \centering
 \includegraphics[scale=.45]{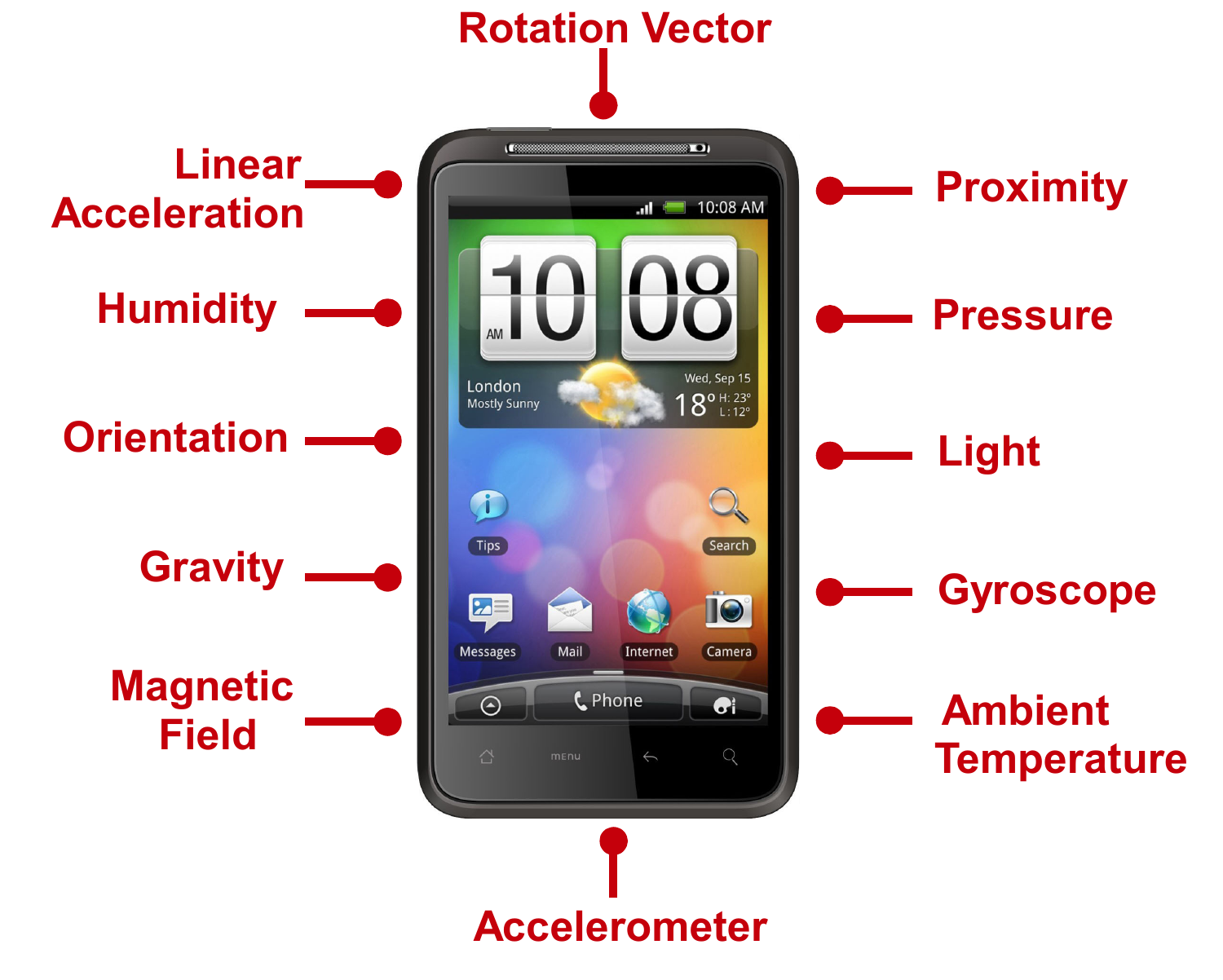}
\vspace{-0.23cm}	
 \caption{Sensors in Mobile Phones}
 \label{Sensors in Mobile Phones}	

\vspace{-0.23cm}	
\end{figure}

In this research, we focus on the Android platform and Android enabled mobile phones. Most modern mobile phones have variety of different sensors built into them. These sensors can be divided into three main categories: \textit{motion sensors} (accelerometer, gravity, gyroscope, linear accelerometer, and rotation vector), \textit{position sensors} (orientation, geomagnetic field, and proximity), and \textit{environment sensors} (light, pressure, humidity and temperature). Figure \ref{Sensors in Mobile Phones} shows the sensors supported by smart phones based on the Android platform 4.0.


Even though, the Android platform supports twelve different sensors, the mobile phones may not have all these sensors built into them. Sensors built into a mobile phone could be varied depending on the hardware manufacturer. There are few other sensors built into the mobile phones: microphone as the audio sensor, camera as the video sensor, digital compass, and Global Positioning System(GPS) sensors. In this research, we do not consider these types of sensors. The details of the sensors supported by the latest Android platform 4.0 are available at \cite{P224}. A description and the unit of measure are also provided for each raw sensor readings.

Even though the sensing capability of a mobile phone is limited to few sensors, it can be easily extended by using Personal Area Network (PAN) technologies such as IrDA, Bluetooth, Wireless USB, Z-Wave, Near field communication (NFC) and ZigBee\footnote{IrDA [irda.org]; Bluetooth [bluetooth.com]; Wireless USB [usb.org]; Z-Wave [www.z-wave.com]; Near field communication [nfc-forum.org]; ZigBee [zigbee.org]}. PAN technologies can connect additional sensing devices to the mobile phones.

Access to the sensors built into the mobile phones to retrieve raw sensor readings are the challenges that we have addressed. We propose to connect mobile phones to a data stream processing engine. We use Global Sensor Network (GSN) middleware as our data stream processing engine to accomplish this task. The advantage to connecting mobile phones to a data stream processing engine is that, raw sensor data can be queried and manipulated with other sensors connected to the engine. GSN allows us to build virtual sensors that can be used to combine verity of different sensors together to create complex outputs. The next section briefly describes the Global Sensor Network middleware.


\section{Global Sensor Network}
\label{sec:Global Sensor Networks}

The Global Sensor Network (GSN) \cite{P022,P050} is a platform aimed at providing flexible middleware to address the challenges of sensor data integration and distributed query processing. It is a generic data stream processing engine. GSN has gone beyond the traditional sensor network research efforts such as routing, data aggregation, and energy optimisation. The design of GSN is based on four basic principles: simplicity, adaptivity, scalability, and light-weight implementation. GSN middleware simplifies the procedure of connecting heterogeneous sensor devices to applications. Specifically, GSN provides the capability to integrate, discover, combine, query, and filter sensor data through a declarative XML-based language and enables zero-programming deployment and management. The above reasons lead us to choose GSN as our data processing engine over other alternative solutions.

The GSN is based on a container based architecture. A detailed explanation is provided in \cite{P022}. The \textit{Virtual Sensor} is the key element in the GSN. A virtual sensor can be any kind of data producer, for example, a real sensor, a wireless camera, a desktop computer, a mobile phone, or any combination of virtual sensors. Typical, a virtual sensor can have multiple input data streams but have only one output data stream.


A \textit{Wrapper} is a piece of Java code that does the data acquisition for a specific type of device. The GSN is capable of retrieving data from various data sources. Wrappers are used to accomplish this task. Wrappers transform the raw data into the GSN standard data model that can be queried and manipulated later. All the wrapper classes need to extend the \textit{AbstractWrapper} class. Typically, third party libraries are initialised in the wrapper constructor. Each sensor needs to have a specific wrapper that can  be used to retrieve raw sensor data. In order to connect a Mica2 \cite{P152} sensor, for example the GSN should have a corresponding wrapper that can talk to Mica2 sensors and retrieve data from it. Currently, the GSN provides a wrapper for all TinyOS \cite{P149} based sensors. Likewise, in order to connect Android phone's built-in sensors to the GSN, it has to have a wrapper that can retrieve raw sensor data from Android phones. \textit{Android Wrapper} which we developed for GSN is around 400 line of code. We discuss GSN wrappers in general and wrapper's life cycle in details in the Section \ref{sec:GSN Wrapper's Life Cycle}. We also provide a detailed explanation about the \textit{Android Wrapper} in Section \ref{sec:Android Wrapper}. The GSN is an open source middleware platform and implementation is available in \cite{P227}.

\section{DAM4GSN Architecture}
\label{sec:System Description}
We discuss the proposed Data Acquisition Model For GSN (DAM4GSN) architecture by dividing it into three separate sections: server configuration, client configuration, and data formats. The server configuration section explains how the GSN server needs to be configured in order to collect sensor data from mobile phones. The client configuration section explains how the mobile phone needs to be configured in order to read the sensor data through built-in sensors and send them to the GSN server. The data format section explains how the communication between GSN server and mobile phone can be done and how the data packets are formatted.

\begin{figure}[!ht]
 \centering
 \includegraphics[scale=.45]{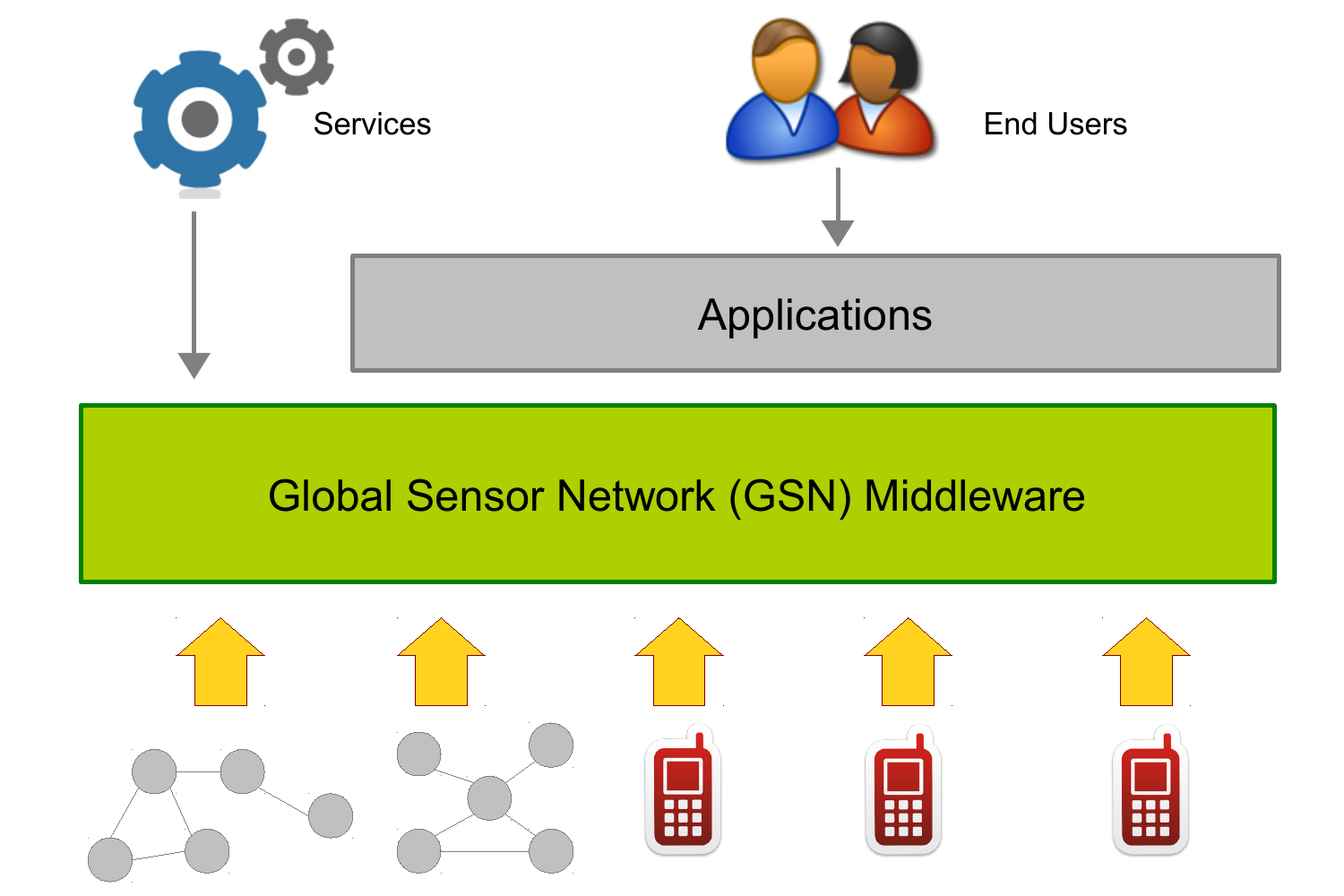}
\vspace{-0.23cm}	
 \caption{Collect Data through Mobile Phones}
 \label{Collect Data through Mobile Phones}	
 \vspace{-0.23cm}
\end{figure}

The overall system architecture is depicted in Figure \ref{Collect Data through Mobile Phones}. The GSN middleware gathers raw sensor data from mobile phones and organizes them according to the GSN standard data model and sends data to applications or services when requested.

\subsection{Server Level Configuration}
\label{sec:Server Level Configuration}

\begin{figure}[!ht]
 \centering
 \includegraphics[scale=1]{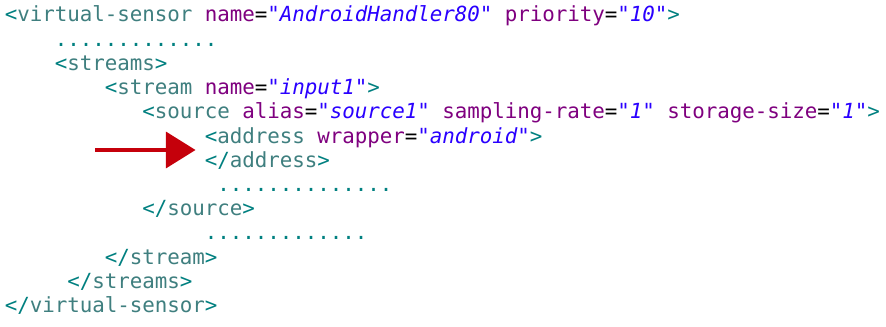}
 \vspace{-0.73cm}	
 \caption{Virtual Sensor Definition File}
 \label{Virtual Sensor}	
 \vspace{-0.23cm}	
\end{figure}

The GSN server configuration that needs to be done in order to collect sensor data from mobile phones is twofold. First, a wrapper needs to be developed in order to retrieve data from mobile devices. The second step is to define a virtual sensor. Defining a virtual sensor in the GSN server is beyond the scope of this paper. A detailed description about defining a virtual sensor is provided in \cite{P022, P227, P167}. However, we provide a segment of a virtual sensor definition in Figure \ref{Virtual Sensor}. A Virtual Sensor Definition (VSD) file provides the information to the GSN that is required to create a Virtual Sensor. This definition is an XML file that contains a predefined set of elements. Here, we only focus on the address element as pointed out in the Figure \ref{Virtual Sensor}.

This is the element that tells the GSN which wrapper to use to retrieve sensor data. Based on this virtual sensor definition, a virtual sensor will be created and an \textit{AndroidWrapper} will be initialised and used to retrieve sensor data to the GSN server.

The communication between the GSN server and client application in the mobile phone can be done through a wrapper. Therefore, we developed a GSN wrapper to accomplish this task. The communication between the wrapper and mobile device is based on client-server architecture.


The communication has two phases. Phase (1) communication happens only once where the client sends metadata to the GSN server. Then, the GSN server configures the wrapper based on the metadata and gets ready to accept the raw sensor data which is constructed by the client using the same metadata. 

After the connection has been established, the client will start sending sensor data to the GSN server according the preconfigured frequency. We developed a generic wrapper that can accommodate any type of sensor reading. For example, one phone may be configured to sense only temperature. In another instance, another phone may be configured to sense temperature, humidity, pressure and accelerometer values. Our wrapper can handle any of these configurations. The wrapper can dynamically change its internal data structures to facilitate any sensing scenario. A detailed description on wrapper development is available in \cite{P167, P227, P022}. The GSN server procedure can be explained as follows.


%

\begin{program}

\textbf{Server Procedure:}
Input: List Of Client Connections (C) = \{ c_{1} \dots c_{n} \} 

Read  The Virtual Sensor Definition();
Wrapper \longleftarrow Identify the Matching Wrapper (VSD);
Virtual Sensor \longleftarrow Create The Virtual Sensor (Wrapper);

\FOR i:=1 \TO size(C) \STEP 1 \DO

    c_{i} \longleftarrow  C \{ c_{1} \dots c_{n} \};
    connection \longleftarrow isClientsFirstConnection(c_{i});
    
    \DO \IF connection;
    metaData \longleftarrow getMetaData(c_{i});
    createDataStructure (metaData);
   \ELSE
    sensorData \longleftarrow getSensorData(c_{i});
    mapSensorDataToGSNDataModel(sensorData);

\END
\END
\END
\end{program}


\subsection{Client Level Configuration}
\label{sec:Client Level Configuration}

\begin{figure}[h]
 \centering
 \includegraphics[scale=.40]{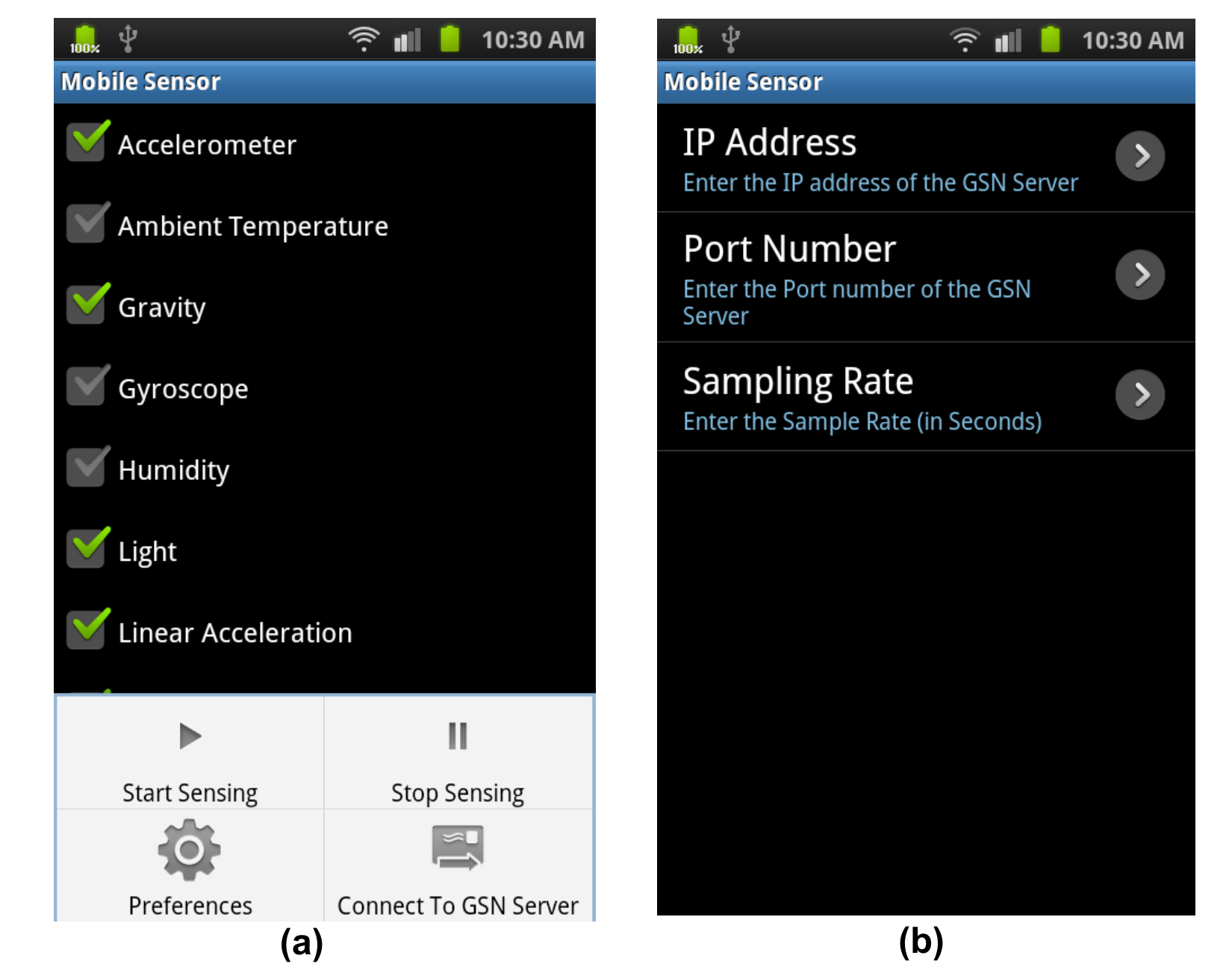}
\vspace{-0.23cm}	
 \caption{Client Application}
 \label{Client Application}	
\vspace{-0.53cm}	
\end{figure}

Mobile users use different versions of the Android operating systems in their hand held devices \cite{P224}. Therefore, we built the client application to support all different platforms. When the application starts running, it automatically identifies the sensors built into the mobile phone hardware. Each sensor will be enabled only if it is supported by both hardware and software layers as shown in Figure \ref{Client Application} (a). Then, the users can select the required sensors that they wants to be used for sensing. After selecting the sensors, the client application requires the GSN server IP address, the port number, and sensing frequency to be configured in the preferences screen as shown in the Figure \ref{Client Application} (b). After configuration is done, the user needs to connect the device to the selected GSN server via WiFi or 3G. 

The connection will be established based on the metadata packet sent to the GSN server by the client application. The details of the metadata packet is discussed in Section \ref{sec:Network Protocol}. After a connection between the GSN server and mobile application is established, the user can press \textit{`Start Sensing'} in the options menu to start the sensing service. The client application will generate and send sensor data packets based on the selected sensors and the sensing frequency to the GSN server. The format of the sensor data packet is discussed in Section \ref{sec:Network Protocol}.

\begin{program}
\textbf{Client Procedure:}
Input: List Of Selected Sensors (S) = \{ S_{1} \dots S_{n} \} 
Output: sensorDataPacket\\ 
\newline

Identify Supported Sensors();
S = Let The User To Select Sensors ();
metaData = Generate the MetaData Packet (S);

connection \longleftarrow Connect To GSN Server(metaData);

\IF connection;

\WHILE (User Stop Sensing)
sensorData = Generate SensorData Packet (S);
Send Data Packet To GSN Server(sensorData)

\END
\END
\end{program}


\subsection{Data Format}
\label{sec:Network Protocol}

\begin{figure}[!ht]
 \centering
 \includegraphics[scale=.50]{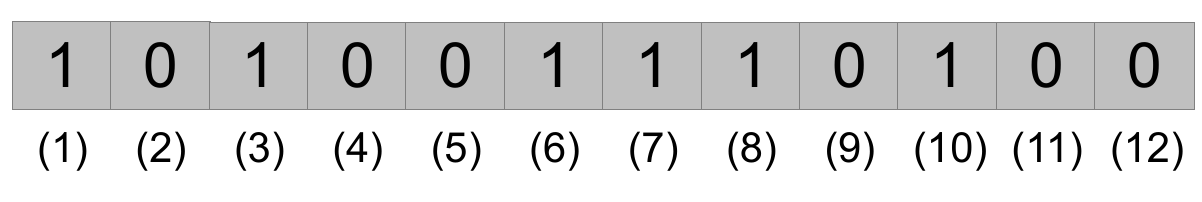}
\vspace{-0.33cm}	
 \caption{Metadata Packet}
 \label{Metadata Packet}	
\vspace{-0.33cm}	
\end{figure}

We define a data packet format for the communication between GSN server and the mobile phone. Our proposed format consists of two parts: metadata and sensor data. A Metadata packet is used to establish the connection between the mobile phone and the GSN server. Figure \ref{Metadata Packet} shows a sample metadata packet. It consists of twelve Boolean values.

%
%
%

Each index corresponds to a sensor in the mobile phone: accelerometer (1), gravity (2),gyroscope (3), linear acceleration  (4), rotation vector (5), magnetic field (6),orientation (7), proximity (8),temperature (9),light (10), pressure (11), humidity (12). The Boolean value of the each index is true if the mobile device is configured to use the corresponding sensor. For example, Figure \ref{Metadata Packet} shows a sample metadata packet where the mobile phone is configured to sense using  Accelerometer Sensor, Gravity Sensor, Linear Acceleration Sensor, Rotation Vector Sensor, Magnetic Field Sensor, Orientation Sensor, Orientation Sensor, Proximity Sensor, and Light Sensor. Only the indexes correspond to above sensors have the value true. The total size of the metadata packet is 12 bits. The metadata packet size does not vary from devices to device. It is a fixed size packet.

A full sensor data packet includes all 27 floating point sensor readings. In contrast to the metadata packet, length of the sensor data packet can be varied. The length of the packet can be varied from four bytes (one float value) to 108 bytes (27 float values). The client application generates sensor data packet in the same order as the selected sensors in the metadata packet. Let's consider the same configuration as previously explained in metadata. 

The client application checks whether the first sensor, accelerometer sensor, is configured to be sensed. If the correspondent value returns true, then the sensor readings related to  accelerometer will be appended to the sensor data packet. Like wise all the sensors are validated and values are appended to the sensor data packet and finally sends to the GSN server. GSN server can correctly understand the sensor data packet based on the metadata packet in similar way.

\section{GSN Wrapper's Life Cycle}
\label{sec:GSN Wrapper's Life Cycle}

The life cycle of a wrapper, depicted in Figure \ref{Life Cycle of Android Wrapper}, begins with the virtual sensor definition (VSD). When a user defines a VSD file, it triggers the virtual sensor creation processes. This process triggers the specified wrapper to be created. The wrapper that correspond to each stream source is defined under the address element in a VSD file. This process sends a Wrapper Connection Request (WCR) to the wrapper repository in the GSN server. A Wrapper Connection Request is an object which contains a wrapper name and its initialisation parameters as defined in the Virtual Sensor. Whenever a WCR is generated at the virtual sensor loader, it will be sent to the wrapper repository. Then the following steps are followed:

\begin{figure}[h]
 \centering
 \includegraphics[scale=.95]{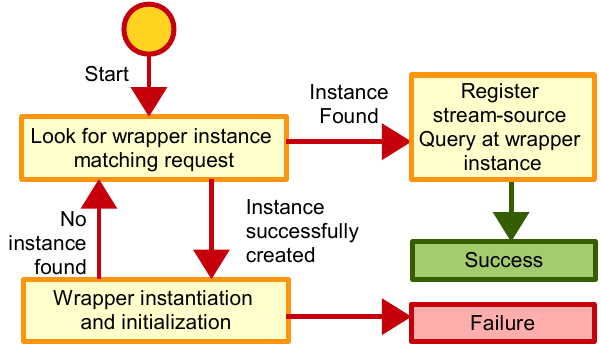}
\vspace{-0.23cm}	
 \caption{Life Cycle of Android Wrapper}
 \label{Life Cycle of Android Wrapper}	
\vspace{-0.33cm}	
\end{figure}

\begin{itemize}
\item First, wrapper repository looks for a wrapper instance that matches to the wrapper connection request. If found, then the stream-source query will be registered with the wrapper and returns true. Sensor data will be captured based on the registered query.
\item If there aren't any WCR in the repository, the wrapper repository creates a new appropriate wrapper object. Then, the newly created object would be added to the wrapper repository. Finally, the stream-source query will be registered with the wrapper and returns true.
\item If there aren't any WCR in the repository and wrapper repository does not have an appropriate wrapper it returns false. The virtual sensor loader fails to load a virtual sensor if at least one of the stream sources required by an input stream fails. For example, if the user define a virtual sensor as depicted in the Figure \ref{Virtual Sensor} and if the GSN wrapper repository does not have an Android wrapper, then the virtual sensor would fail.
\end{itemize}

\section{Android Wrapper}
\label{sec:Android Wrapper}

\begin{figure}[h]
 \centering
 \includegraphics[scale=.9]{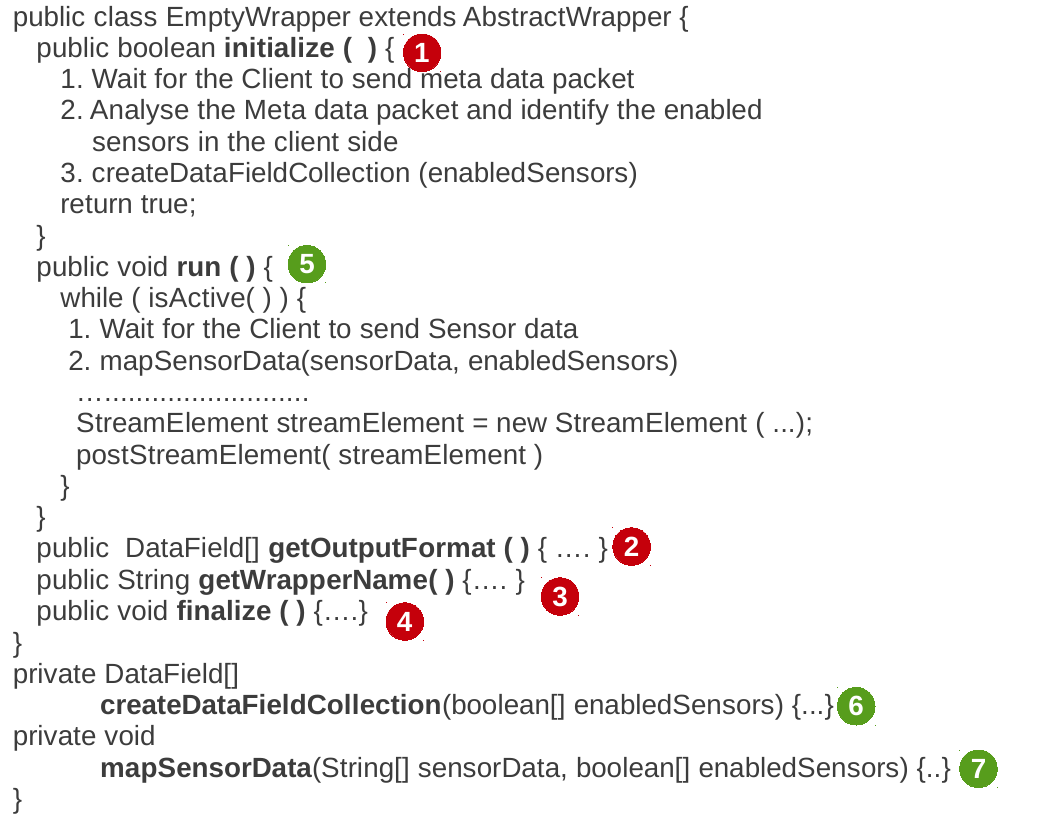}
\vspace{-0.83cm}	
 \caption{Android Wrapper}
 \label{Android Wrapper}	
\vspace{-0.23cm}	
\end{figure}

In this section we discuss the Android wrapper we created. We provide the details about how to create a wrapper according to the GSN standards. A code segment of the Android wrapper is presented in Figure \ref{Android Wrapper}. Our objective of the code demonstration is to explain the few methods that are essential in wrapper creation. The depicted code can be explained as a template for wrapper creation.

All the wrappers need to extend the Java class\newline \textit{gsn.wrapper.AbstractWrapper}. Therefore, all the wrappers are  subclasses of AbstractWrapper. There are four methods that need to be implemented by the subclasses. Those methods are numbered 1-4 in the Figure \ref{Android Wrapper}. The methods are 1. \textit{boolean initialise()}, 2. \textit{void finalise()}, 3. \textit{String getWrapperName()}, 4. \textit{DataField[] getOutputFormat()}

A new thread is created for each wrapper in GSN. After creating the wrapper object, \textit{initialise()} method is called as shown in number (1). All the communication using third party libraries should happen within this method. For example, a camera wrapper may talk to a third party API in order to talk to the camera and retrieve the camera images. In the Android wrapper we developed, \textit{initialise()} method creates a socket and waits until the client mobile phone send the metadata packet. Metadata packet is required to determine the data structure. Once the meta packet is received, it is analysed and identified the client side sensing capability. This will determine the sensors available in the client mobile phone. This information is passed into the \textit{createDataFieldCollection()} as shown in number (6).

The method \textit{finalise()} is called at the end of the wrappers life cycle. This is the last chance you get to release all the resources. This method can be used to close all the connections we established with the outer world. Concretely, all the resources acquired during the \textit{initialise()} method should be released here. For example, if you open a file in the initialisation phase, you should close it in the finalisation phase. In the Android wrapper, all the client communication resources such as sockets and ports are released in this method.

The method \textit{getWrapperName()} returns the name of the wrapper. The method \textit{ getOutputFormat()} returns a \textit{DataField} object that provides a description of the data structure produced by the wrapper.

The \textit{run()} method is responsible for retrieving sensor data from sensors and forwarding them to the GSN middleware. In the Andorid wrapper, \textit{run()} method waits until the mobile phone clients send the sensor data. Once the data is received, \textit{mapSensorData()} method maps the newly received data to the GSN data model structure which is created in the wrapper initialisation phase. However, the packet format of the metadata and sensor data packet need to be matched. After the data has been inserted into the GSN data model, querying, filtering and other functionalities provided by GSN can be done over the mobile phone sensor data. 


%
%
%
%
%

\begin{figure*}
 \centering
 \includegraphics[scale=.45]{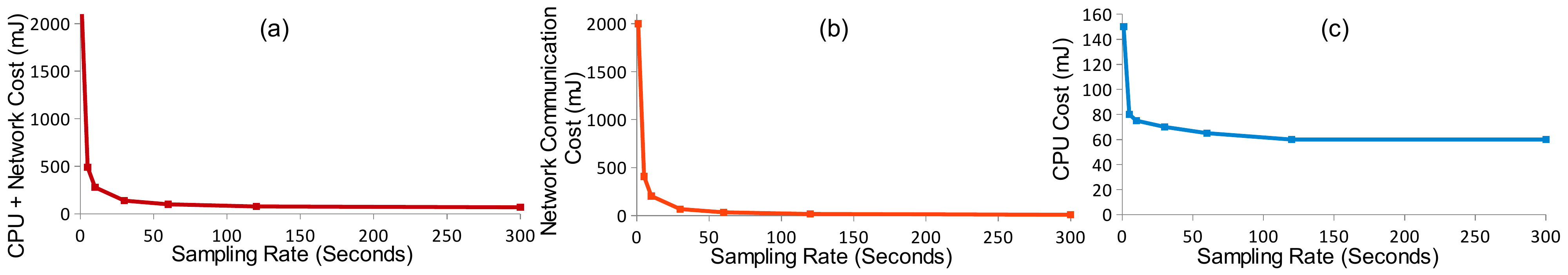}
\vspace{-0.63cm}	
 \caption{Performance Evaluation. All the sensors available in the mobile phone are used for the evaluation. Energy consumption in mJ per minute}
 \label{Performance Evaluation}	
\vspace{-0.53cm}	
\end{figure*}

\section{Performance Advantage}
\label{sec:Performance Benefits}
There are several advantages in our approach over the existing GSN communication model.

\begin{itemize}
\item GSN assumes that sensors are connected to a server that is running GSN middleware. However, installing and configuring GSN in many computers would be a overwhelming task. GSN middleware requires certain amount of processing power \cite{P050} as well. In contrast, our approach demonstrated how the sensor data can be captured through a client device (in our scenario, a mobile phone) and send them to the GSN server over a network. In GSN side we only need to develop a single wrapper that can transform sensor data packets into GSN model as demonstrated in the \textit{AndroidWrapper}. As we do not port (install) GSN into  mobile devices, scalability is preserved at the server level, probably in the cloud.

\item Any form of update may only be required to be done in the client side (i.e in mobile phones or computers). No update is required in GSN server. In contrast, if we connect a sensor such as SunSPOT to the GSN server via wired connection, then we have to develop a separate SunSPOT wrapper for GSN, which we may need to manually develop it, compile it and then attach it to the GSN server. Any requirement of update may need to perform the above process repetitively in the GSN server.

\item Sensing capability of the mobile phones can be extended by attaching additional hardware components. It is not required to do any changes in wrappers in GSN server.

\item Our approach can be used by any mobile device or low end computing devices (e.g. iPhone, iPad, Windows phone, Blackberry, etc.). The only capability that a mobile device need to have is sensor packet generation and network communication. These functionalities are commonly available in modern devices.

\end{itemize}

\section{Evaluation}
\label{sec:Evaluation}

We conducted all evaluations and experiments using a \textit{Samsung Galaxy S} mobile phone which runs Android platform 2.3 and PowerTutor\footnote{ziyang.eecs.umich.edu/projects/powertutor} app. GSN instance was installed on a laptop with Intel Core i7 CPU and 6GB ram. Network communications are through CSIRO ict center wifi network. 
Each sensor consumes different amount of power. Accelerometer, gravity, linear acceleration sensors consume 0.20 mA (each). Further, proximity and light sensors consume 0.75 mA (each). Magnetic field sensor consumes 4.00 mA. In addition, linear acceleration and orientation sensors consume 4.20 mA (each). According to above information, rotation vector sensor. magnetic field sensor and orientation sensor consume 21 times more power than accelerometer, gravity, linear acceleration sensor and 5.6 times more power than proximity and light sensors.

The three graphs depicted in Figure \ref{Performance Evaluation} shows how the energy cost varies with the sampling rate. According to the graph, network communication cost is always higher than the CPU energy cost. Therefore, network communication parameters such as sampling rate should be carefully planned in order to utilise the sensors built into the mobile phones in an efficient manner. Further, data compression techniques can be employed to reduce the amount of data to be sent to the GSN. Further evaluation on energy consumption based on data packet size, distance, sampling rate is presented in \cite{P510}.

\section{Conclusion}
\label{sec:Conclusion}

In this research work, we identified how the sensor data can be captured using mobile phones. In our approach, we evaluate the process of connecting a sensor to a data processing engine called GSN. \textit{AndroidWrapper} was developed in order to retrieve sensor data from mobile phones. DAM4GSN allows GSN to collect sensor data from low-level computational devices such as mobile phones without porting GSN itself into mobile phones. However, it was realised that each and every low-level sensor, that does not have computational capabilities, should have a wrapper talking to a GSN server in order to collect data. 

Developing such wrappers is a time consuming and tedious job. Therefore, in future, we intend to conduct our research towards automating wrapper development. The proposed DAM4GSN architecture will be built into the GSN middleware in the future releases.


%
%



%
%

\bibliography{Bibliography}
\bibliographystyle{abbrv}

\end{document}